\documentstyle[11pt,aaspp4,flushrt]{article}
 
\newcommand{\apg}{\gtrsim}
\newcommand{\apl}{\lesssim}

\newcommand{\etal}{et al.}

\newcommand{\kms}{\mbox{km\ s${^{-1}}$}}
\newcommand{\lya}{\mbox{${\rm Ly}\alpha$}}

\begin{document}
 
\lefthead{Chen et al.}
 
\righthead{}

%\slugcomment{Submitted to The Astrophysical Journals}
 
\title{THE ORIGIN OF \lya\ ABSORPTION SYSTEMS AT $z>1$---IMPLICATIONS FROM 
THE HUBBLE DEEP FIELD}

\author{HSIAO-WEN CHEN\altaffilmark{1} and KENNETH M. LANZETTA}
\affil{Department of Physics and Astronomy, State University of New York at
Stony Brook \\
Stony Brook, NY 11794--3800, U.S.A. \\
hchen,lanzetta@sbastr.ess.sunysb.edu}

\altaffiltext{1}{Current address: The Observatories of the Carnegie Institution
of Washington, 813 Santa Barbara Street, Pasadena, CA 91101, U.S.A.\\
hchen@ociw.edu} 

\author{ALBERTO FERN\'{A}NDEZ-SOTO\altaffilmark{2}}
\affil{Department of Astrophysics and Optics, School of Physics, University of 
New South Wales \\
Kensington--Sydney, NSW 2052, AUSTRALIA \\
fsoto@bat.phys.unsw.edu.au}

\altaffiltext{2}{Current address: Department of Physics and Astronomy, State 
University of New York at Stony Brook \\
Stony Brook, NY 11794--3800, U.S.A. \\
fsoto@sbastr.ess.sunysb.edu}

%\newpage
 
\begin{abstract}

    The Hubble Deep Field images have provided us with a unique chance to 
relate statistical properties of high-redshift galaxies to statistical 
properties of \lya\ absorption systems. Combining an {\em empirical} measure of
the galaxy surface density versus redshift with an {\em empirical} measure of 
the gaseous extent of galaxies, we predict the number density of \lya\ 
absorption systems that originate in extended gaseous envelopes of galaxies 
versus redshift.  We show that at least 50\% and as much as 100\% of 
observed \lya\ absorption systems of $W\apg0.32$ \AA\ can be explained by 
extended gaseous envelops of galaxies. Therefore, we conclude that
known galaxies of known gaseous extent must produce a significant fraction and 
perhaps all of \lya\ absorption systems over a large redshift range. 

\end{abstract}
 
 
\newpage
 
\section{INTRODUCTION}
 
    Comparison of galaxy and absorber redshifts along common lines of sight 
demonstrates that a significant and perhaps dominant fraction of low-redshift 
($z < 1$) \lya-forest absorption systems arise in extended gaseous envelopes 
of galaxies (Lanzetta \etal\ 1995a; Chen \etal\ 1998).  But it has not yet been
possible to extend this comparison to higher redshifts, because normal surveys 
fail to identify galaxies at redshifts much beyond $z = 1$.  Nevertheless,
determining the origin of high-redshift \lya-forest absorption systems bears 
crucially on all efforts to apply the absorbers as a means to study galaxies 
in the very distant universe and to study how the extended gas of galaxies 
evolves with redshift. Previous arguments based on the apparent low metal 
content and lack of clustering of the absorbers suggested an intergalactic 
origin (Sargent \etal\ 1980), but these arguments are weakened by recent Keck 
measurements of the actual metal content and clustering of the absorbers 
(Songaila \& Cowie \etal\ 1996; Fern\'{a}ndez-Soto \etal\ 1996).

    To extend the comparison of galaxies and absorbers to higher redshifts, it 
is necessary to systematically identify normal galaxies at redshifts $z>1$. 
Over the past few years, it has become clear that high-redshift galaxies can 
be reliably identified by means of broad-band photometric techniques. For 
example, several groups have determined photometric redshifts of galaxies in 
the Hubble Deep Field (HDF) to redshifts as large as $z \approx 6$ (Lanzetta, 
Yahil, \& Fern\'andez-Soto 1996; Sawicki, Lin, \& Yee 1997; Connolly \etal\ 
1997; Lanzetta, Fern\'andez-Soto, \& Yahil 1998; Fern\'andez-Soto, Lanzetta, \&
Yahil 1999). Spectroscopic redshifts of nearly 120 of these galaxies obtained 
using the Keck telescope (see Fern\'andez-Soto \etal\ 1999 for a complete list
of references) 
%Cohen \etal\ 1996; Cowie 1997; Steidel \etal\ 1996; Zepf 
%\etal\ 1997; Lowenthal \etal\ 1997; Dey \etal\ 1998; Weymann \etal\ 1998a; 
%Spinrad \etal\ 1998) 
have demonstrated that the photometric redshifts are both reliable enough and 
accurate enough to establish galaxy surface density versus redshift to large 
redshifts (Lanzetta \etal\ 1998; Fern\'andez-Soto \etal\ 1999).

    Here we combine an {\em empirical} measure of the galaxy surface density
versus redshift, obtained from the HDF, with an {\em empirical} measure of the
gaseous extent of galaxies, obtained by Chen \etal\ (1998), to predict the 
number density of \lya\ absorption systems that originate in extended gaseous 
envelopes of galaxies versus redshift. Previous comparison of galaxies and 
\lya\ absorption systems at redshifts $z < 1$ showed that (1) galaxies of all 
morphological types possess extended gaseous envelopes and (2) the gaseous 
extent of galaxies scales with galaxy $B$-band luminosity but does not depend 
sensitively on redshift (Chen \etal\ 1998). If these results apply to galaxies 
at all redshifts, then known galaxies of known gaseous extent must produce 
some fraction of \lya\ absorption systems at all redshifts. 

    On the basis of our analysis, we find that this fraction is significant. 
Specifically, considering \lya\ absorption systems of absorption equivalent 
width $W\apg0.32$ \AA, we find that galaxies can account for nearly all 
observed \lya\ absorption systems at $z<2$ and that galaxies of luminosity 
$L_B \apg 0.05 L_{B_*}$ can account for approximately 50\% of the observed 
\lya\ absorption systems at higher redshifts. We further argue that if the 
gaseous extent of galaxies does not decrease with increasing redshift, then 
known galaxies must produce {\em at least} as many \lya\ absorption systems as 
our predictions. We show that we can already explain 70\% and perhaps all of 
the observed \lya\ absorption systems at $z>2.0$ after correcting for faint 
galaxies that are below the detection threshold of the HDF images. 

    We adopt a Hubble constant, $H_0=100\ h$ \kms\ Mpc$^{-1}$, and a 
deceleration parameter, $q_0=0.5$, throughout the paper, unless otherwise 
noted.

\section{METHODS}

    To calculate the predicted number density of \lya\ absorption systems that
originate in extended gaseous envelopes of galaxies versus redshift, we 
multiply the galaxy number density per unit comoving volume by the absorption 
gas cross section of galaxies, which is known to scale with properties of 
galaxies.  Our previous comparison of galaxies and \lya\ absorption systems at 
$z<1$ showed that the amount of gas encountered along a line of sight depends 
on galaxy impact parameter $\rho$ and $B$-band luminosity $L_B$ but does not 
depend strongly on galaxy average surface brightness $\langle \mu \rangle$, 
disk-to-bulge ratio $D/B$, or redshift $z$ (Chen \etal\ 1998). Namely, galaxies
of all morphological types possess extended gaseous envelopes, and the 
absorption gas cross section scales with galaxy $B$-band luminosity but does 
not evolve significantly with redshift. The left panel of Figure 1 shows a
statistically significant correlation between galaxy $B$-band magnitude and the
residuals of the $W$ versus $\rho$ anti-correlation. In contrast, the right 
panel of Figure 1 shows no correlation at all between galaxy redshift and the 
residuals of the $W$ versus $\rho$ anti-correlation accounting for galaxy 
$B$-band luminosity. 

    The predicted number density of \lya\ absorption systems $n(z)$ originating
in extended gaseous envelopes of galaxies can now be written
\begin{equation}
n(z) = \frac{c}{H_0} (1 + z) (1 + 2 q_0 z)^{-1/2} 
\int^\infty_{L_{B_{\rm min}}} dL_B \ \Phi(L_B,z) \pi R^2(L_B),
\end{equation}
where $c$ is the speed of light, $\Phi(L_B,z)$ is the galaxy luminosity 
function, and $R$ is the gaseous extent. The minimum galaxy $B$-band luminosity
$L_{B_{\rm min}}$ is determined by the detection threshold of either the galaxy
redshift survey or the \lya\ absorbing galaxy survey. It is taken to be the 
brighter of the magnitude limits at which the galaxy luminosity function or the
scaling relation was determined. Supplementing results of Chen \etal\ (1998) 
with new measurements (Chen \etal, in preparation), we find that galaxy gaseous
radius $R$ scales with galaxy $B$-band luminosity $L_B$ as
\begin{equation}
\frac{R}{R_*}=\left(\frac{L_B}{L_{B_*}}\right)^{t},
\end{equation}
with
\begin{equation}
t=0.39\pm0.09,
\end{equation}
and with $R_* = 172\pm 27 \ h^{-1} \ {\rm kpc}$ at $W=0.32$ \AA\ for galaxies 
of luminosity $L_B > 0.03 L_{B_*}$. 

    Given a complete galaxy sample, we can evaluate equation (1) by 
establishing an empirical galaxy luminosity function without adopting any
particular functional form. First, we write the galaxy luminosity function as 
a discrete sum of $\delta$ functions,
\begin{equation}
\Phi(L_B,z)=\left(\frac{c}{H_0}\right)^{-1}(1+z)^{-1}(1+2q_0z)^{1/2}
\frac{1}{\Delta z\ \Omega\ D_A(z)^2}\sum_i \delta(L_B-L_{B_i}),
\end{equation}
where $\Omega$ is the angular survey area, $D_A$ is the angular diameter 
distance, and $L_{B_i}$ is the $B$-band luminosity of galaxy $i$. Next, we
substitute equation (4) into equation (1) and rearrange terms to yield
\begin{equation}
n(z) = \frac{1}{\Omega}\frac{1}{(z_2 - z_1)}\sum_i 
\frac{\pi R^2(L_{B_i})}{D_A^2(z_i)},
\end{equation}
where $(z_1,z_2)$ marks the boundary of each redshift bin. Equation (5) 
indicates that the number density of \lya\ absorption systems originating in 
the extended gaseous envelopes of galaxies is equal to the product of the 
galaxy surface density and the average of luminosity weighted gas cross 
sections. 

\section{GALAXY SAMPLE}

    Here we adopt a galaxy sample from the Hubble Deep Field (HDF) galaxy 
catalog published by Fern\'{a}ndez-Soto \etal\ (1999), which contains 
coordinates, optical and near-infrared photometry, and photometric redshift 
measurements of 1067 galaxies. This galaxy catalog is unique for studying 
galaxy statistics at high redshifts, because it is sensitive to objects of 
surface brightness down to $\approx 26.1\ {\rm mag}\ {\rm arcsec}^{-2}$ and is 
complete to $AB(8140) = 28.0$ within the central $3.93\ {\rm arcmin}^2$ area of
the HDF image (Zone 1 of Fern\'{a}ndez-Soto \etal\ 1999). Photometric redshifts
ranging through $z\approx 6$ were measured from broad-band spectral energy 
distributions established from the optical (Williams \etal\ 1996) and infrared 
(Dickinson \etal\ 1999, in preparation) images. Spectroscopic redshifts of 
nearly 120 of these galaxies have been obtained using the Keck telescope (see 
Fern\'andez-Soto \etal\ 1999 for a complete list of references).  Comparison of
the photometric and spectroscopic redshifts shows that the photometric 
redshifts are accurate to within an RMS relative uncertainty of $\Delta
z/(1 + z) < 0.1$ at all redshifts $z < 6$ (Lanzetta \etal\ 1998;
Fern\'andez-Soto \etal\ 1999), except that there are not spectroscopic 
redshifts available for comparison at redshifts $z=1.5$ to $z=2.0$.  Reliable 
galaxy statistics as a function of redshift can therefore be measured on the 
basis of the HDF galaxy catalog. 

\section{ANALYSIS}

    The goal of the analysis is to compare the predicted number density of 
\lya\ absorption systems originating in extended gaseous envelopes of galaxies 
(using the HDF galaxy catalog) with the observed number density of \lya\ 
absorption systems.

    Here we evaluate equation (5) for galaxies observed in the HDF, given the
photometric redshift measurements. Galaxy $B$-band luminosity is calculated 
from the apparent F814W magnitude corrected for luminosity distance, color 
$k$ correction (between rest-frame $B$ band and observed-frame F814W band, 
calculated based on the spectral type determined from the photometric redshift 
techniques), and bandpass $k$ correction. Errors in the predicted number 
density are estimated using a standard bootstrap method, which incorporates the
sampling error with the uncertainties of photometric redshift measurements and 
the scaling relation. To model the uncertainty of photometric redshift 
measurements correctly, Lanzetta \etal\ (1998) have shown that both photometric
error and cosmic variance with respect to the spectral templates must be 
accounted for. The first can be simulated by perturbing galaxy photometry of 
different bands within the photometric errors. The second can be characterized 
by the RMS dispersion of photometric and spectroscopic redshift measurements, 
which is 0.08 at $z<2$ and 0.32 at  $z>2$ (Lanzetta \etal\ 1998). The 
uncertainty of photometric redshift measurements is then calculated by forming 
a quadratic sum of the two.

    The results are shown in Table 1 and Figure 2 in comparison with 
observations. In Figure 2, filled circles represent the observations and open 
circles represent our predictions. The observations at $z<1.5$ are taken from
the sample 2 of Weymann \etal\ (1998), which we normalize to refer to 
\lya\ absorption lines of $W\apg0.32$ \AA. The observations at $z>1.6$ are 
taken from the sample 9 of Bechtold (1994)\footnote{Note that the data from 
Weymann \etal\ included all the \lya\ absorption lines with or without 
associated metal absorption lines, while the data from Bechtold included only 
the \lya\ absorption lines without associated metal absorption lines. However,
the correction for including metal-line associated systems in the high-redshift
sample is expected to be approximately a few percent (Bechtold 1994; Frye 
\etal\ 1993). In comparison to the large uncertainty of the number density of
high-redshift \lya\ absorption systems, it is clear that the absence of 
metal-line associated systems in the high-redshift sample does not change the
redshift distribution of \lya\ absorption systems by a noticeable amount.}.  
We select a bin size, $\Delta z=0.5$, when calculating the predicted values. 
Experiments with different bin sizes show that the redshift distribution of the
predicted number density of \lya\ absorption systems is insensitive to the 
selected bin size.  We also repeat the calculation, using a deceleration 
parameter, $q_0=0.0$. It turns out that the predicted number density of \lya\ 
absorption systems does not depend sensitively on the adopted deceleration 
parameter either.

    Our predictions have so far been limited to including galaxies of 
luminosity $L_B > L_{B_{\rm min}}$. To estimate the contribution of faint 
galaxies to the predicted gas cross section, we calculate the number density of
\lya\ absorption systems originating in galaxies of luminosity $L_B < L_{B_{\rm
min}}$ by choosing a Schechter luminosity function. We evaluate equation (1) by
adopting a faint-end slope obtained by Ellis \etal\ (1996), which is $\alpha = 
-1.41_{-0.07}^{+0.12}$, and by extrapolating the scaling relation (equation 2 
and 3) to obtain the extended gaseous radii for galaxies of luminosity $0\apl 
L_B < 0.03 L_{B_*}$. Increments of the predicted number density versus redshift
due to the inclusion of faint galaxies are shown in Figure 2 as well (crosses 
with dashed horizontal bars). 

\section{DISCUSSION}

    The HDF images have provided us with a unique chance to relate statistical
properties of high-redshift galaxies to statistical properties of \lya\ 
absorption systems, thereby studying the origin of \lya\ absorption systems at 
high redshifts. Combining an {\em empirical} measure of galaxy surface density 
versus redshift with an {\em empirical} measure of gaseous extent of galaxies, 
we have predicted the number density of \lya\ absorption systems that originate
in extended gaseous envelopes of galaxies versus redshift.

    For \lya\ absorption systems of absorption equivalent width $W\apg0.32$ 
\AA, comparison of the predicted and observed number densities of \lya\ 
absorption systems shows that (1) known galaxies with known gas cross 
sections can account for all the observed \lya\ absorption systems to within 
measurement errors at redshifts $0<z<1$, (2) known galaxies of luminosity
$L_B \apg 0.03\ L_{B_*}$ with unevolved gas cross sections can account for all 
\lya\ absorption systems at redshifts $1<z<2$, and (3) known galaxies of 
luminosity $L_B \apg 0.05\ L_{B_*}$ with unevolved gas cross sections can 
account for approximately 50\% of observed \lya\ absorption systems at $z>2$. 
Apparently, bright galaxies alone can account for a significant portion of the 
observed \lya\ absorption systems at all redshifts. After correcting for faint
galaxies that are below the detection threshold of the HDF image, we can 
already explain 70\% and perhaps all of the observed \lya\ absorption systems 
at $z>2.0$ to within errors. Here we discuss three factors that may invalidate
the result.

    First, we consider evolution of the galaxy luminosity function. There has 
been some evidence showing that galaxies at the faint end of galaxy luminosity 
function evolves significantly with redshift (e.g. Ellis 1997, and references 
therein). Namely, faint galaxies may be more numerous at higher redshifts. To 
address this issue, we evaluate equation (1) with $L_{B_{\rm min}}\approx 
0$ to calculate the relative change in the predicted number density, given a 
different faint-end slope. It turns out that the predicted number density of 
\lya\ absorption systems increases by as much as $ 0.3$ dex, if we vary the 
faint-end slope from $\alpha=-1.4$ to $\alpha=-1.6$. Apparently, a steeper
faint-end slope brings the predicted values closer to the observed ones. 
We conclude that known galaxies must produce {\em at least} as many \lya\ 
absorption systems as the predictions.

    Next, we consider evolution of neutral gas surrounding galaxies. In the 
right panel of Figure 1, we show that the residuals of the $W$ versus $\rho$ 
anti-correlation after accounting for the scaling of galaxy $B$-band 
luminosity $L_B$ do not correlate with galaxy redshift. Although there 
is no direct observational evidence to support the extension of the redshift 
independence of the scaling relation to redshifts beyond $z=1$, the gaseous 
extent of galaxies is unlikely to decrease with increasing redshift in these
early epochs. Theoretically, it is believed that galaxies were formed through 
accretion of cooled gas over time, indicating a larger gaseous extent of 
galaxies at higher redshifts. Observationally, the mass density of neutral gas 
measured from damped \lya\ absorption systems increases significantly with 
increasing redshift at redshifts between $z=1.6$ and $z=3.5$ (Lanzetta, Wolfe,
\& Turnshek 1995; Storrie-Lombardi, McMahon, \& Irwin 1996), supporting that 
galaxies may possess more neutral gas and therefore no less gaseous extent at 
higher redshifts. We again conclude that known galaxies must produce {\em at 
least} as many \lya\ absorption systems as the predictions.

    Finally, we consider galaxy clustering effects. Due to limited resolutions 
of spectroscopic observations, \lya\ absorption lines with small velocity 
separation are sometimes blended together and considered as one absorption 
system. Lanzetta, Webb, \& Barcons (1996) first reported an absorption system 
which may arise in a group or cluster of galaxies. It was later confirmed by
Ortiz-G\'{\i}l \etal\ (1999), who identified eight absorption features in this
system based on a spectrum of higher spectral resolution. Their analysis 
suggested that the degree of clustering of \lya\ absorption systems may be 
underestimated on velocity scales of several hundred kilometers per second (see
also Fern\'andez-Soto \etal\ 1996). Similarly, the number density of \lya\ 
absorption lines derived from a known galaxy population would be in excess of 
the observed number density of \lya\ absorption lines, because galaxies are
strongly clustered and the predicted number density does not suffer from the
line blending effect. As a result, the fraction of \lya\ absorption systems 
originating in extended gaseous envelopes of galaxies may be overestimated. 

   To estimate the amount of excess at different redshifts, we calculate the 
expected number of neighbouring galaxies of luminosity $L_B>L_{B_{\rm min}}$, 
$N_{\rm neighbours}$, that are seen at impact parameters $\rho\apl\ 200$ kpc 
along a line of sight with a velocity difference of $\Delta v$ from an 
absorption redshift.  We adopt a two-point correlation function measured by 
Magliocchetti \& Maddox (1999) and Arnouts \etal\ (1999) for the HDF galaxies 
at redshifts $0\apl z\apl 4.8$, which is characterized by 
\begin{equation}
\xi(r,z)=(r/r_0)^{-\gamma}(1+z)^{-(3+\epsilon)}
\end{equation}
with $\gamma=3+\epsilon=1.8$ and $r_0\approx 1.7\ h^{-1}$ Mpc, and evaluate the
integral of the two-point correlation function over a comoving volume spanned 
by $\delta v$ in depth and 200 kpc in radius. The velocity span, $\Delta v$, is
taken to be the spectral resolution of which the observations were carried out.
It is $250\ \kms$ for the sample from Weymann \etal\ (i.e.\ absorption systems 
at $z<1.5$) and $75\ \kms$ for the sample from Bechtold (i.e.\ absorption 
systems at $z>1.5$). We show in Table 2 that the estimated excess in our 
predictions (represented by the number of neighbouring galaxies) can be at 
most a factor of four at $z<2$, but is negligible at $z>2$. Therefore, we 
conclude that the fraction of \lya\ absorption systems originating in extended 
gaseous envelopes of galaxies remains significant even after correcting for 
galaxy clustering.

    In summary, we present our first attempt at relating statistical properties
of galaxies to statistical properties of \lya\ absorption systems at high 
redshifts based on the HDF galaxy catalog. Combining an {\em empirical} 
measure of galaxy surface density versus redshift with an {\em empirical} 
measure of gaseous extent of galaxies, we have predicted the number density of 
\lya\ absorption systems that originate in extended gaseous envelopes of 
galaxies versus redshift.  We show that at approximately 50\% and as much as 
100\% of observed \lya\ absorption systems of $W\apg0.32$ \AA\ can be explained
by extended gaseous envelops of galaxies. The result remains valid after 
taking into account possibile evolutions of faint galaxies and extended gas
around galaxies, as well as galaxy clustering effects. Therefore, we conclude 
that known galaxies of known gaseous extent must produce a significant fraction
and perhaps all of \lya\ absorption systems over a large redshift range. 

\acknowledgments
 
  The authors thank John Webb for helpful discussions. HWC and KML were 
supported by NASA grant NAGW--4422 and NSF grant AST--9624216.  AF was 
supported by a grant from the Australian Research Council. 

\newpage
 
\begin{deluxetable}{cccc}
\tablecaption{NUMBER DENSITY OF \lya\ ABSORPTION SYSTEMS WITH $W\apg0.32$ 
\AA---PREDICTED AND OBSERVED}
\tablewidth{0pt}
\tablehead{\colhead{}  & \multicolumn{2}{c}{$n(z)$} & \colhead{} \\
\cline{2-3} \\
\colhead{Redshift}  & \colhead{Predicted} & \colhead{Observed} & 
\colhead{$L_{\rm min}/L_{B_*}$}\tablenotemark{a}}
\startdata  
      $0.0 < z < 0.5$ & 13 & 27 & 0.03 \nl
      $0.5 < z < 1.0$ & 29 & 28 & 0.03 \nl
      $1.0 < z < 1.5$ & 27 & 30 & 0.03 \nl
      $1.5 < z < 2.0$ & 34 & 30 & 0.03 \nl
      $2.0 < z < 2.5$ & 26 & 58 & 0.05 \nl
      $2.5 < z < 3.0$ & 25 & 70 & 0.09 \nl
      $3.0 < z < 3.5$ & 24 & 77 & 0.12 \nl
      $3.5 < z < 4.0$ & 13 & 89 & 0.17 \nl
\tablenotetext{a}{Here we adopt an $L_{B_*}$ corresponding to $M_{B_*} = -19.5$
(Ellis et al.\ 1996).}
%\tablecomments{}
\enddata
\end{deluxetable}

%\begin{deluxetable}{rr}
%\tablecaption{VARIATION OF PREDICTED NUMBER DENSITY WITH FAINT GALAXY COUNTS}
%\tablewidth{0pt} 
%\tablehead{\colhead{$\alpha$}  & \colhead{$\Delta\log n(z)$}  }
%\startdata  
%      $-1.0$ & $-0.291$ \nl
%      $-1.3$ & $-0.108$ \nl
%      $-1.4$ & $ 0.000$ \nl
%      $-1.5$ & $ 0.116$ \nl
%      $-1.6$ & $ 0.289$ \nl
%      $-1.7$ & $ 0.529$ \nl
%\enddata
%\end{deluxetable}
\begin{deluxetable}{cc}
\tablecaption{ESTIMATED EXCESS OF PREDICTED \lya\ ABSORPTION SYSTEMS}
\tablewidth{0pt} 
\tablehead{\colhead{Redshift}  & \colhead{$N_{\rm neighbours}$}  }
\startdata  
      $0.05 < z < 0.5$ & $1.7$ \nl
      $0.5 < z < 1.0$ & $2.7$ \nl
      $1.0 < z < 1.5$ & $3.3$ \nl
      $1.5 < z < 2.0$ & $1.1$ \nl
      $2.0 < z < 2.5$ & $0.8$ \nl
      $2.5 < z < 3.0$ & $0.6$ \nl
      $3.0 < z < 3.5$ & $0.5$ \nl
      $3.5 < z < 4.0$ & $0.4$ \nl
\enddata
\end{deluxetable}

\begin{figure}
\centerline{\vbox{\plottwo{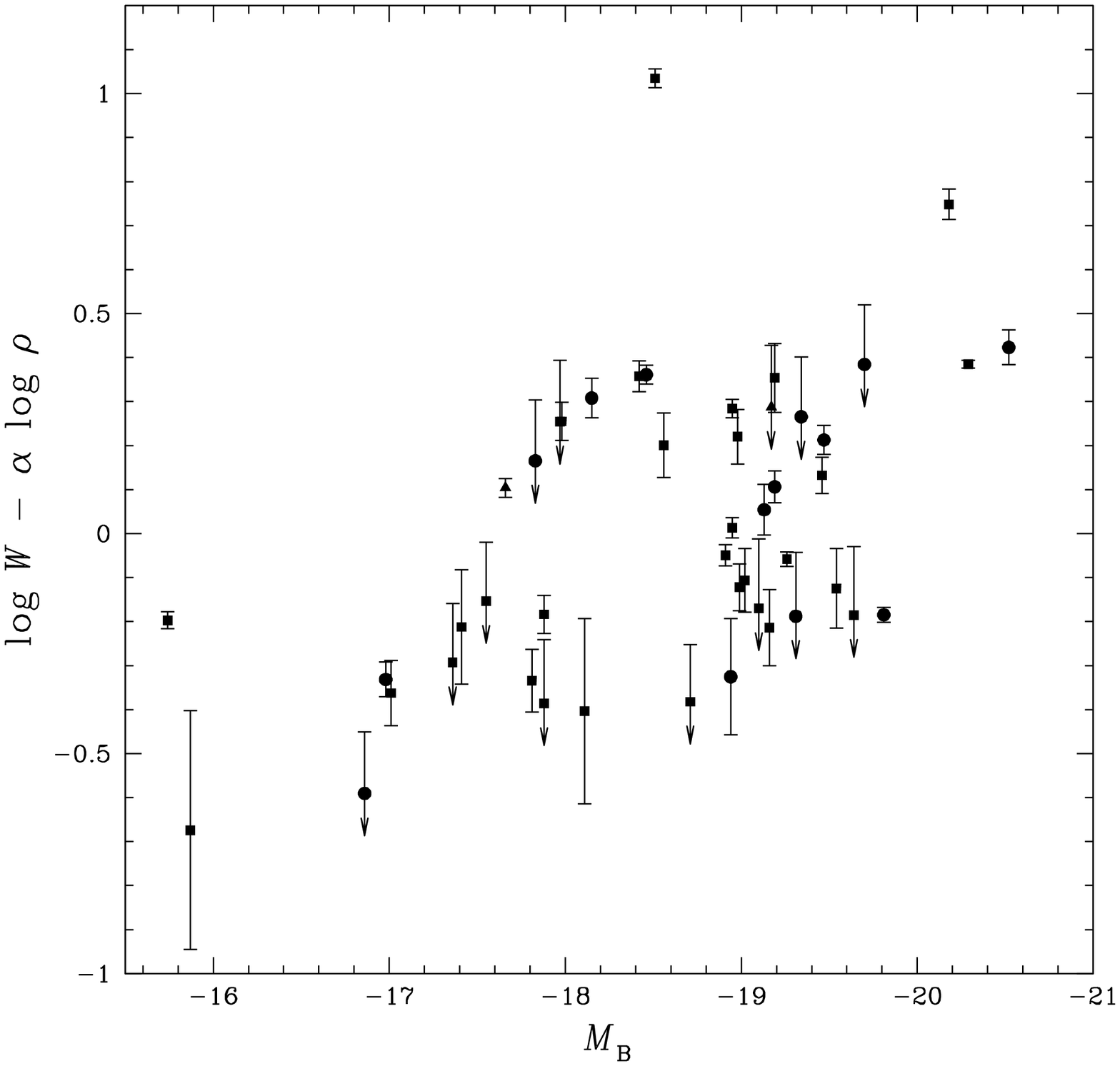}{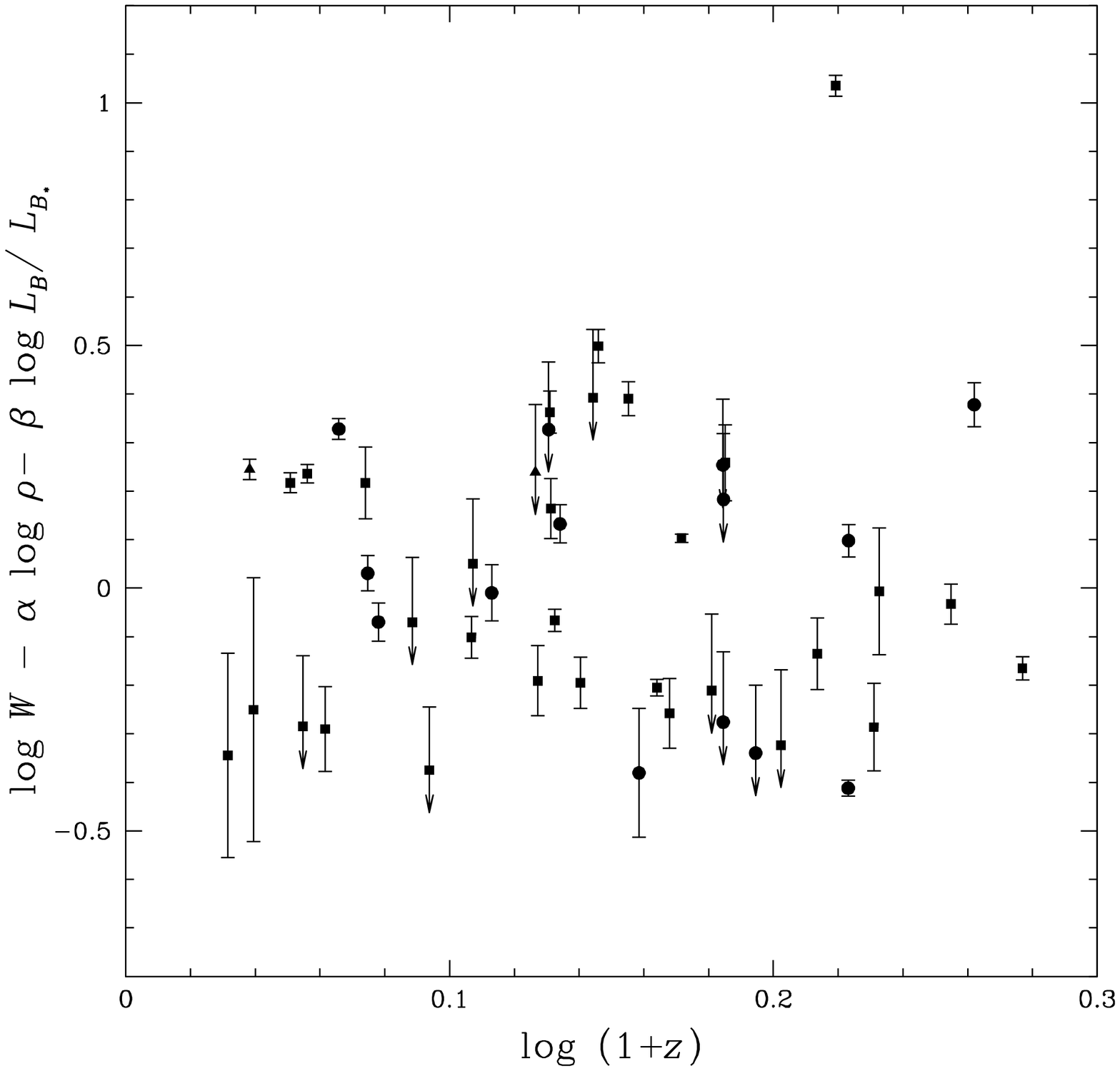}}}
\caption[]{Left: The residuals of the $W$ vs. $\rho$ anti-correlation as a
function of galaxy $B$-band magnitude. Right: The residuals of the $W$ vs. 
$\rho$ anti-correlation accounting for galaxy $B$-band luminosity as a function
of galaxy redshift. Solid circles represent elliptical or S0 galaxies, 
triangles represent early-type disk galaxies, and squares represent late-type 
disk galaxies. Points with arrows indicate no detections at a 3-$\sigma$ upper 
limit. The coefficients, $\alpha$ and $\beta$, were determined from a maximum 
likelihood analysis previously discussed in Chen \etal\ (1998).}
\end{figure}

\newpage

\begin{figure}
\centerline{\vbox{\plotone{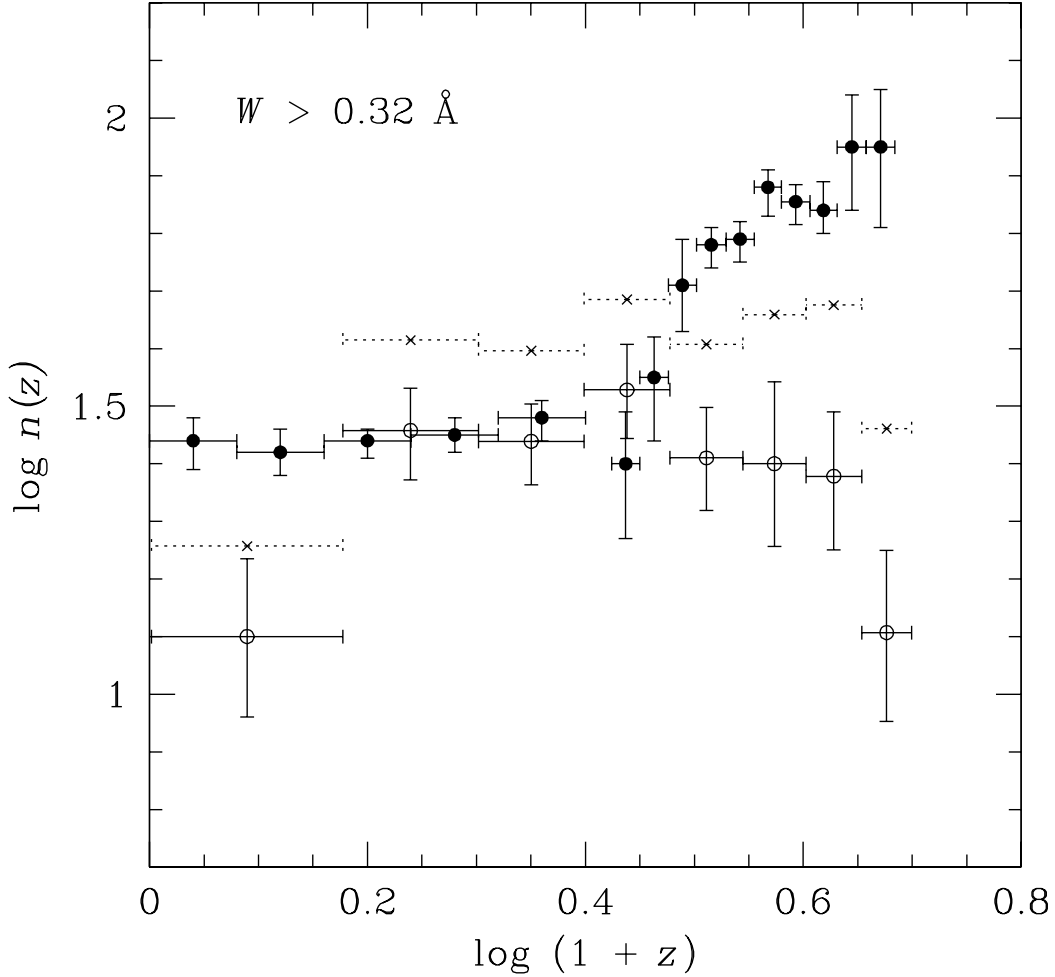}}}
\caption[]{Comparison of the observed number densities of \lya\ absorption
systems of absorption equivalent width $W>0.32$ \AA\ (closed circles) and 
the predicted ones produced in extended gaseous envelopes of galaxies observed 
in HDF (open circles). The absorption line observations are from Bechtold 
(1994) for data at $z>1.6$ and from Weymann \etal\ (1998b) for data at $z<1.5$.
We have multiplied the data from Weymann \etal\ by a factor of
$\exp\left(\frac{0.32-0.24}{0.273}\right)$ in order to be compared with 
observations from Bechtold. The vertical bars represent 1-$\sigma$ errors in 
the observed and predicted values and the horizontal bars indicate the adopted 
bin size, $\Delta z=0.5$. Crosses indicate the predicted number densities 
of \lya\ aborption systems arising in galaxy gaseous envelopes after correcting
for galaxies fainter than the detection limit of the HDF image by taking 
$L_{B_{\rm min}} = 0.0$ and adopting a galaxy luminosity function obtained by 
Ellis \etal\ at $0.15 < z < 0.35$ (1996).}
\end{figure}

\end{document}